\documentclass{llncs}
\usepackage{makeidx}  
\usepackage{cleveref}
\usepackage{graphicx}
\usepackage{amsfonts}
\usepackage{amsmath}
\usepackage{mathtools}
\usepackage{algpseudocode, algorithm}

\newcommand\coolover[2]{\mathrlap{\smash{\overbrace{\phantom{%
    \begin{matrix} #2 \end{matrix}}}^{\mbox{$#1$}}}}#2} 

\newcommand\coolunder[2]{\mathrlap{\smash{\underbrace{\phantom{%
    \begin{matrix} #2 \end{matrix}}}_{\mbox{$#1$}}}}#2}

\newcommand\coolrightbrace[2]{%
\left.\vphantom{\begin{matrix} #1 \end{matrix}}\right\}#2}

\graphicspath{{pics/}{figs/}}

\begin{document}
\frontmatter          
\pagestyle{headings}  
\mainmatter              
\title{Comparison of Brain Networks with Unknown Correspondences}
\titlerunning{Comparison of Brain Networks with Unknown Correspondences}  
\author{Sofia Ira Ktena\thanks{This work has been funded by EPSRC and IPEP (\url{http://ipep-gr.org/}).} \and Sarah Parisot \and Jonathan Passerat-Palmbach  \and Daniel Rueckert}
\authorrunning{Ktena et al.}   
%
\tocauthor{author1 (Affiliation of author1),
author2 (Affiliation of Author2), Daniel Rueckert (Affiliation of
Author3), }

\institute{Biomedical Image Analysis Group, Imperial College London, UK 
}


\maketitle              

\begin{abstract}
Graph theory has drawn a lot of attention in the field of Neuroscience during the last decade, mainly due to the abundance of tools that it provides to explore the interactions of elements in a complex network like the brain. The local and global organization of a brain network can shed light on mechanisms of complex cognitive functions, while disruptions within the network can be linked to neurodevelopmental disorders. In this effort, the construction of a representative brain network for each individual is critical for further analysis. Additionally, graph comparison is an essential step for inference and classification analyses on brain graphs. In this work we explore a method based on graph edit distance for evaluating graph similarity, when correspondences between network elements are unknown due to different underlying subdivisions of the brain. We test this method on 30 unrelated subjects as well as 40 twin pairs and show that this method can accurately reflect the higher similarity between two related networks compared to unrelated ones, while identifying node correspondences.
\end{abstract}



\section{Introduction}
The term human connectome was first coined a decade ago by~\cite{sporns2005human} and~\cite{hagmann2005diffusion} to describe the structural white matter connections, which can be mapped with advanced neuroimaging techniques. Similarly, the putative functional connections between spatially remote regions can be identified with functional neuroimaging or electrophysiological recording techniques~\cite{friston1993functional}. The resulting connectomes comprise complete maps of the brain's structural or functional connections at the macro-scale and embody the notion of representing all connections within the brain as graphs. Analysis of these graphs opens new experimental and theoretical avenues in several areas of neuroscience, since brain connectivity is critical to neurodevelopment, neurodegeneration and the neural basis of cognition~\cite{sporns2011networks,smith2016linking}. Looking at brain connectivity from a network perspective takes advantage of modern graph-theoretic approaches to study the brain and allows the discovery of patterns that emerge as an outcome of highly selective and structured coupling between network elements~\cite{sporns2011networks}.

Since most machine learning techniques, including classifiers, regressors and kernel machines, entail some sort of (dis)similarity measures, the definition of a meaningful similarity measure is of major importance for the inference procedure. Nodes in brain networks represent regions obtained with a certain parcellation technique, which often does not account for variations in anatomy and function of individual brains~\cite{zalesky2010whole}. Individual-based parcellations are therefore being sought for the delineation of boundaries between nodes in a biologically meaningful way - such as functional task-based or connectivity-based parcellations in individual subject's space - which, however, give rise to unknown correspondences. Hence, a desirable property of the similarity measure is to allow comparisons between graphs where node correspondence is not guaranteed.

Methods based on the embedding of graphs to obtain a general vector representation have been widely used~\cite{varoquaux2013learning} to assess graph similarity. However, these methods, also referred to as ``bag of edges''~\cite{craddock2013imaging}, discard all information about the structure of the graph and cannot be applied to graphs with distinct sets of nodes. Another group of methods is based on the notion of similarity between two objects that is evaluated on an implicitly induced feature space \cite{mokhtari2013decoding,jie2014topological}. Most generic graph kernels compare features of small subgraphs extracted from the original graph leading to an inherently local perspective, which may fail to capture global properties of graphs \cite{vishwanathan2010graph}. Alternatively, graph invariants like small-world index, modularity and global efficiency, which have been widely used in neuroscience studies \cite{van2010exploring}, significantly reduce the dimensionality of the graph comparison problem, but are sometimes hard to interpret biologically and discard a great amount of local information. 

We propose to use graph edit distance to evaluate brain graph similarity, since this method is able to match the graphs directly in their domain~\cite{gao2010survey}. This approach is, therefore, able to model structural variation of graphs in a very intuitive and illustrative way, while considering both their structural and semantic information. However, the structural similarity of two graphs is only correctly reflected by graph edit distance, if the underlying costs are defined appropriately, a property that has been disregarded in its limited applications on brain graphs~\cite{raj2010network}. In this work we developed an efficient version of graph edit distance tailored for brain graphs, which takes into account both spatial constraints as well as local network information as a node's ``signature''. We evaluate this method on the structural and functional networks of 30 unrelated healthy subjects and 40 female twin pairs and show that our approach can successfully reflect similarities between corresponding networks.


\section{Methods}

\subsection{Dataset and preprocessing}
The minimally preprocessed dataset was provided by the Human Connectome Project~\cite{glasser2013minimal}. We used the diffusion MRI data of 30 unrelated healthy individuals, as well as the diffusion (dMRI)  and functional (fMRI) data of 40 female twin pairs (20 monozygotic and 20 dizygotic). All subject data were registered to a common MNI space and the cortical surface of each individual brain was extracted and represented as a triangular mesh. Subsequently, a probabilistic tractography algorithm~\cite{behrens2007probabilistic} was used to estimate the neural pathways from several seed points (5000 per vertex of the cortical surface mesh), yielding a number of streamlines connecting the seed point with the rest of the vertices. The fMRI data were also preprocessed to remove spatial artifacts and distortions and were, finally, converted to a standard ``grayordinate'' space to facilitate cross-subject comparisons.

\subsection{Network construction}
\begin{figure}[t]
\centering
\includegraphics[width=.88\linewidth]{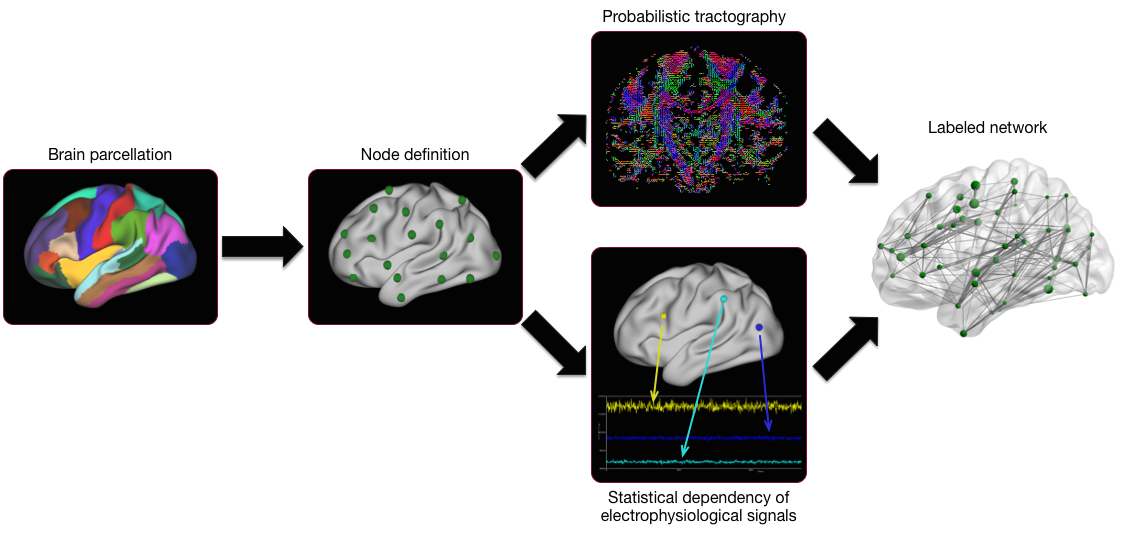}
\caption{\textit{Network construction from diffusion or functional MRI data}. A brain parcellation is required to specify the network nodes. Each node then represents a single region from the parcellation. Probabilistic tractography is used to create a map of white matter connections from dMRI images. The number of streamlines (for dMRI) or statistical dependency of representative timeseries (for fMRI) is used to estimate the structural and functional weights, respectively, yielding a labeled brain network.}
\label{fig:network_construction}
\end{figure}

An undirected, labeled graph $G \in \mathcal{G}$ can be described with a tuple $G = (\mathcal{V}, \mathcal{E}, \mathcal{\mu}, \mathcal{\nu})$, where $\mathcal{V}$ is the finite set of nodes, $\mathcal{E} \subseteq \mathcal{V} \times \mathcal{V}$ is the set of edges, $\mathcal{\mu} : \mathcal{V} \to \mathcal{L_V}$ is the node labeling function and $\mathcal{\nu} : \mathcal{E} \to \mathcal{L_E}$ is the edge labeling function. It should be noted that a \textit{graph} is a formal mathematical representation of a \textit{network}, but the two terms can be used interchangeably in this context. An overview of the steps required to construct a brain network is illustrated in Figure~\ref{fig:network_construction}.

\textbf{Nodes in brain space.} In this work, the nodes are defined using connectivity-driven parcellations performed in each individual's brain space and represent contiguous sets of voxels~\cite{parisot2016grampa}. Data-driven parcellations aim at defining functionally coherent regions based on their structural or functional connectivity profile. The number of parcels $q$ may vary, allowing the construction of networks at different resolutions. In this work we choose $q=50$. The incorporation of functional information in the parcellation is more likely to result in more homogeneous regions that better represent connectivity.
 
\textbf{Edge weights.} We define the edge weights of structural networks as the number of streamlines connecting two regions. Two nodes $u$ and $v$ are connected with an edge $e=(u, v)$ if there is at least one streamline with endpoints in $ROI(u)$ and $ROI(v)$, where $ROI(u)$ is the parcel associated with a node $u$ and $S(u)$ its cortical surface (i.e. number of vertices within the parcel). The weight 

\begin{equation}
w_{uv} = \frac{2}{S(u) + S(v)} \sum_{f \in F_{e}} log(n(f))
\end{equation}

captures the connection density between the end-nodes $u$ and $v$, where $F_{e}$ is the set of all fibers connecting the two regions and $n(f)$ the number of streamlines connecting two voxels from $ROI(u)$ and $ROI(v)$. The sum $S(u) + S(v)$ corrects for the variable size of cortical ROIs represented by the network nodes~\cite{hagmann2008mapping}, while the $log$ reduces the bias against longer streamlines~\cite{jbabdi2009multiple}.

For the functional networks the mean timeseries of each ROI is used as the representative timeseries, while partial correlation is chosen to define the edge weights. Partial correlation discards the indirect connections that are preserved by its prevalent competitor, Pearson's correlation, and is known to be less prone to noise. It can be computed from the inverse of the empirical covariance matrix, ${\bf P}=\Sigma^{-1}$, as $w_{uv}=-{\bf P}_{uv}({\bf P}_{uu}{\bf P}_{vv})^{-1/2}$. 

\subsection{Assessing similarity}
\textit{Graph edit distance (GED)} is the minimum-weight sequence of edit operations required to transform one graph into another and is defined directly in their domain $\mathcal{G}$ as a non-negative function $d_{GED}:\mathcal{G} \times \mathcal{G} \to \mathbb{R}^+$. The basic edit operations that we consider valid for both nodes and edges are: (1) \textit{substitution}: $u \to v$, with $u \in \mathcal{V}_1$ and $v \in \mathcal{V}_2$, (2) \textit{deletion}: $u \to \varepsilon$ and (3) \textit{insertion}: $\varepsilon \to v$. The edit distance between $G_1$ and $G_2$ is then defined as

\begin{equation}
d_{GED}(G_1, G_2) = \min_{e \in E(G_1, G_2)} \sum_i c(e_i)
\end{equation}

where $c(e_i)$ is the cost of an edit sequence from $G_1$ to $G_2$ and $E$ the finite set of edit sequences from $G_1$ to $G_2$. In order to better reflect differences between networks, both structural and \textit{egonet}-based features are incorporated in the edit costs. A node's egonet is the induced subgraph of its neighbouring nodes, hence the egonet-based features provide an accurate signature of each node by means of its connections to the rest of the network.

\textbf{Node edit operations.} The cost of node substitution, $c_{i,j}$, consists of a spatial and a feature component. The spatial component is calculated as the Euclidean distance of the node coordinates (coordinates of the representative voxel in MNI space) normalized by the diameter of the MNI transformed sphere mesh, $\delta$, using $d_{euclidean}(u_P, u_Q) = \frac{\Vert P - Q \Vert}{\delta}$. The egonet-based features taken into account for the feature component are the mean and standard deviation of the degree, strength and clustering coefficient of the neighbouring nodes.
The strength of a vertex $u$ is given by $S_{u} = \sum_{v} w_{uv}$, where the sum is over all edges attached to a node. The local clustering coefficient for an undirected graph is given by:

\begin{equation}
c_u = \frac{1}{deg(u)(deg(u)-1))} \sum_{vw} (w_{uv} w_{uw} w_{vw})^{1/3}
\end{equation}

The feature component is then calculated using the \textit{Canberra distance}~\cite{lance1967mixed}, which is given by $d_{canberra}(u_P, u_Q) = \sum_{i=1}^d \frac{|P_i - Q_i|}{|P_i| + |Q_i|}$ for \textit{d}-dimensional feature vectors of nodes $u_P$ and $u_Q$. Canberra distance is known to be very discriminative due to its sensitivity  to small changes near zero, as well as the normalization of the absolute difference of individual comparisons.

The spatial component of the insertion cost, $c_{\varepsilon,i}$, and  deletion cost, $c_{i, \varepsilon}$, is 1. The feature component corresponds to the betweenness centrality of a node in the network. Betweenness centrality is given by:

\begin{equation}
g_u= \sum_{s \neq u \neq t}\frac{\sigma_{st}(u)}{\sigma_{st}}
\label{eq:betweenness}
\end{equation}

where $\sigma_{st}$ is the number of shortest paths from node $s$ to node $t$ while $\sigma_{st}(u)$ is the number of shortest paths through $u$. This reflects the influence that a certain node has on the network, since it is assumed that information transfer follows the shortest paths.

\textbf{Edge edit operations.} The edge edit cost of an addition/deletion is 1, whereas the edit cost of a substitution is equal to the Canberra distance between the edge weights. The latter serves the purpose of normalizing the cost of absolute difference of edge weights.

The time and space complexity of an exact GED computation is exponential in the number of nodes involved, making computation very expensive for brain graphs, whose size is in the range of hundreds or thousands of nodes. In our framework, graph edit distance is normalised so that $d_{GED} \in [0,1]$, where $d_{GED}=0$ indicates that the two graphs are identical.

\begin{figure}[t]
\centering
\includegraphics[width=.9\linewidth]{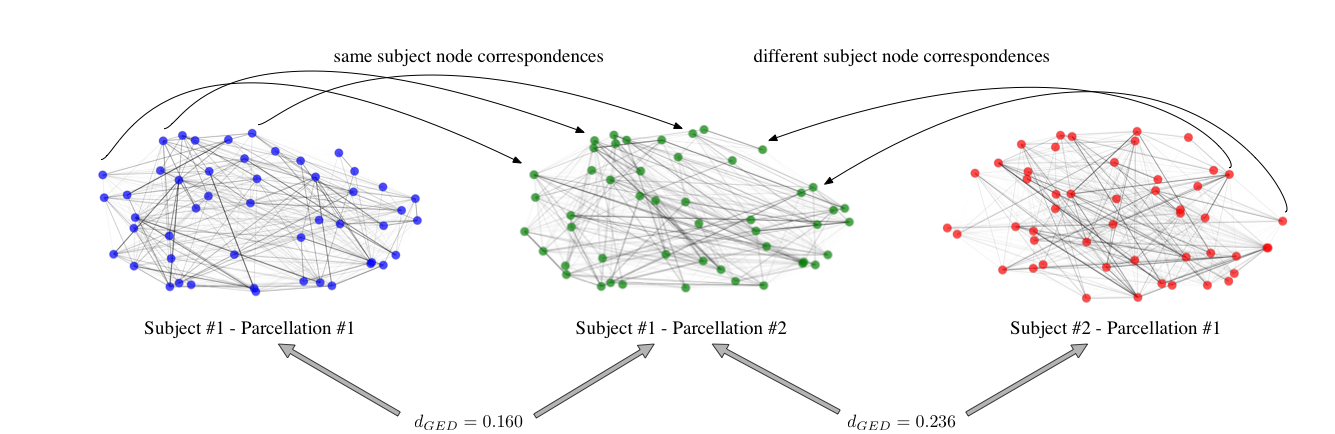}
\caption{\textit{Estimating graph similarity between different networks}. The first two networks belong to the same subject with a different underlying parcellation, whereas the third network represents connectivity for a different subject using the same parcellation as the first one. It can be observed that the distance between the networks of the same subject is lower than the distance between networks from different subjects.}
\label{fig:graph_comparison}
\end{figure}

\subsection{Assignment problem}
As mentioned previously, it is very costly to calculate the exact GED, since its complexity is exponential to the number of nodes. However, the Hungarian algorithm, presented in \cite{riesen2007bipartite}, can provide a fast, locally optimal solution to the exact GED computation. This algorithm is known to solve the assignment problem in cubic time and find the lowest cost assignment between objects from two different sets (see Fig.~\ref{fig:graph_comparison}). The same algorithm can be used to calculate a local optimum solution to the exact GED, by constructing the following cost matrix of order $n+m$, where $|\mathcal{V}(G_1)|=n$ and $|\mathcal{V}(G_2)|=m$:


\[ \vphantom{
    \begin{matrix}
    \overbrace{XYZ}^{\mbox{$substitutions$}}\\ \\ \\ \\ \\ \\ 
    \underbrace{pqr}_{\mbox{$insertions$}}
    \end{matrix}}%
\textbf{C} = \begin{bmatrix}
\coolover{substitutions}{c_{1,1} & \dots & c_{1,m}} & \coolover{}{c_{1,\varepsilon} & \dots & \infty}\\
\vdots & \ddots & \vdots & \vdots & \ddots & \vdots \\
c_{n,1} & \dots & c_{n,m} & \infty & \dots & c_{n,\varepsilon} \\
c_{\varepsilon,1} & \dots & \infty & 0 & \dots & 0 \\
\vdots & \ddots & \vdots & \vdots & \ddots & \vdots \\
\coolunder{insertions}{\infty & \dots & c_{\varepsilon,m}}  & \coolunder{}{0 & \dots &   0 }
\end{bmatrix}%
\begin{matrix}
\coolrightbrace{\infty \\ \vdots \\ c_{\varepsilon,m}}{deletions}\\
\coolrightbrace{0 \\ \vdots \\ 0}{zeros}
\end{matrix}\]

Nevertheless, the above cost matrix does not take the edge edit operations into account. These need to be considered to achieve a better approximation of true edit distance. This is accomplished by adding to each node substitution $c_{i,j}$ the minimum sum of edge edit operation costs implied by this substitution and, similarly, to each node insertion (deletion) the cost of all insertions (deletions) of adjacent edges. The off-diagonal elements of the insertion and deletion cost sub-matrices are set to infinity, since a node can be inserted/deleted only once. The computational complexity of this method is $O((n+m)^3)$.

\subsection{Tailoring GED for brain graphs}
The aforementioned computations are applicable to any kind of graphs, and explore combinations that should be excluded in the special case of brain graphs. Tailoring the GED computation for brain graphs would require the introduction of spatial constraints. Hence, substitutions, which inherently identify node correspondences, should be penalized between nodes that are spatially remote or lie in different hemispheres. In the current framework, only the \textit{n} spatially nearest neighbours of graph $G_2$ are considered for substitution by a node in $G_1$, significantly reducing the computational time. Additionally, a weighting factor $\alpha$ is introduced, which can be tuned depending on the application, in order to control the effect of spatial and feature distance on the total edit distance.

\begin{algorithm}
\begin{algorithmic}
\For{$i=1:n$}
	\For{$j=1:m$}
    	\If{$u_j \in neigh(v_i)$}
        	\State {$c_{i,j} = \alpha * d_{euclidean}(v_i, u_j) + (1-\alpha) * d_{canberra}(v_i, u_j) + d_{edge}(v_i, u_j)$}
        \Else 
        	\State {$c_{i,j} = \infty$}
        \EndIf
    \EndFor
    \State {$c_{i,\epsilon} = \alpha + (1-\alpha) * g(v_i)$}
\EndFor
\For{$j=1:m$}
	\State {$c_{\epsilon,j} = \alpha + (1-\alpha) * g(u_j)$}
\EndFor
\end{algorithmic}
\caption{Construction of cost matrix for GED calculation}
\label{alg:cost_matrix}
\end{algorithm}

In Algorithm~\ref{alg:cost_matrix}, $d_{edge}(v_i, u_j)$ is the edge edit distance between the edges of node $v_i$ from $G_1$ and node $u_j$ from $G_2$, and $g(\cdot)$ is given by Eq.~\ref{eq:betweenness}.

\section{Results}
The proposed method for brain graph comparison was tested in two different settings. First, the method was applied on structural networks generated from different single-subject parcellations to test whether within-subject similarities are reflected with GED. The method was also applied on a dataset of 20 monozygotic (MZ) and 20 dizygotic (DZ) pairs of female twins to explore within-pair distances in comparison to distances between unrelated pairs of subjects.

\begin{figure}[t]
\centering
\includegraphics[width=.85\linewidth]{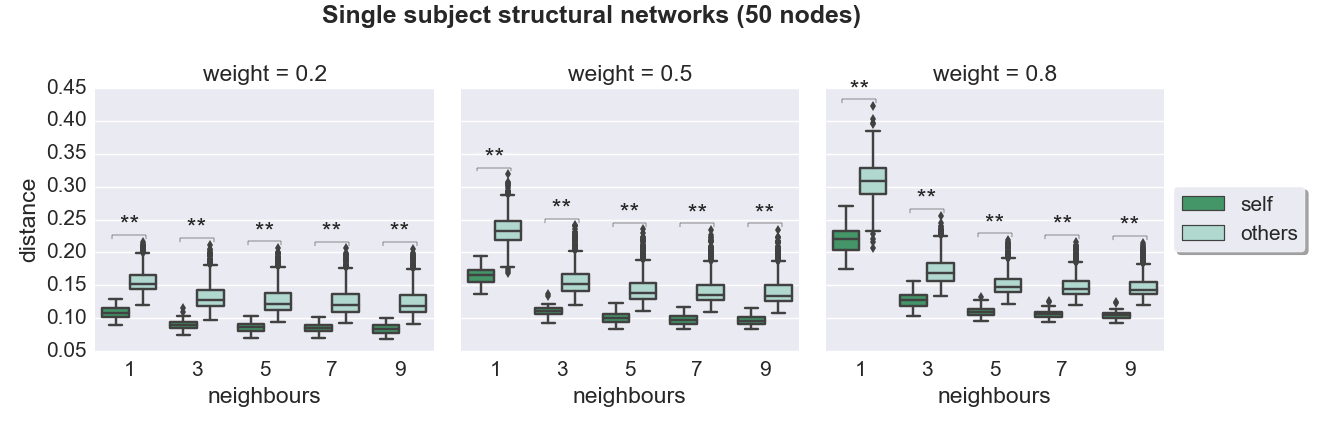}
\caption{Boxplots for $\alpha \in [0.2, 0.5, 0.8]$ comparing within-subject (dark green) to between-subject (light green) GED on structural networks derived from individual parcellations. Permutation tests are performed and statistically significant differences are indicated ($p < 0.05$*, $p < 0.001$**).}
\label{fig:self_vs_others}
\end{figure}

\subsection{Single-subject parcellations}
Two different sets of structural networks were generated for 30 healthy unrelated individuals, which arise from different dMRI-driven subdivisions~\cite{parisot2016grampa} of the left hemisphere into 50 parcels. In this setting the impact of the number of neighbours considered for node substitution as well as the spatial weight, $\alpha$, are explored (Fig.~\ref{fig:self_vs_others}). From the explored parameters the best separation between within-subject and cross-subject distances is achieved for $\alpha=0.5$. This indicates that it is important to incorporate both spatial and feature information about each node, in order to obtain a meaningful distance measure for brain networks. Additionally, it can be observed that increasing the number of neighbours considered for substitution leads to an overall decrease in the calculated distances, which is expected since nodes with more similar connectivity profile can be located marginally further from each other. However, increasing the number of neighbours further than a certain value has only a minor impact on the distances (results for $nn =7,9$ are very similar in all cases).

\begin{figure}[t]
\centering
\includegraphics[width=.85\linewidth]{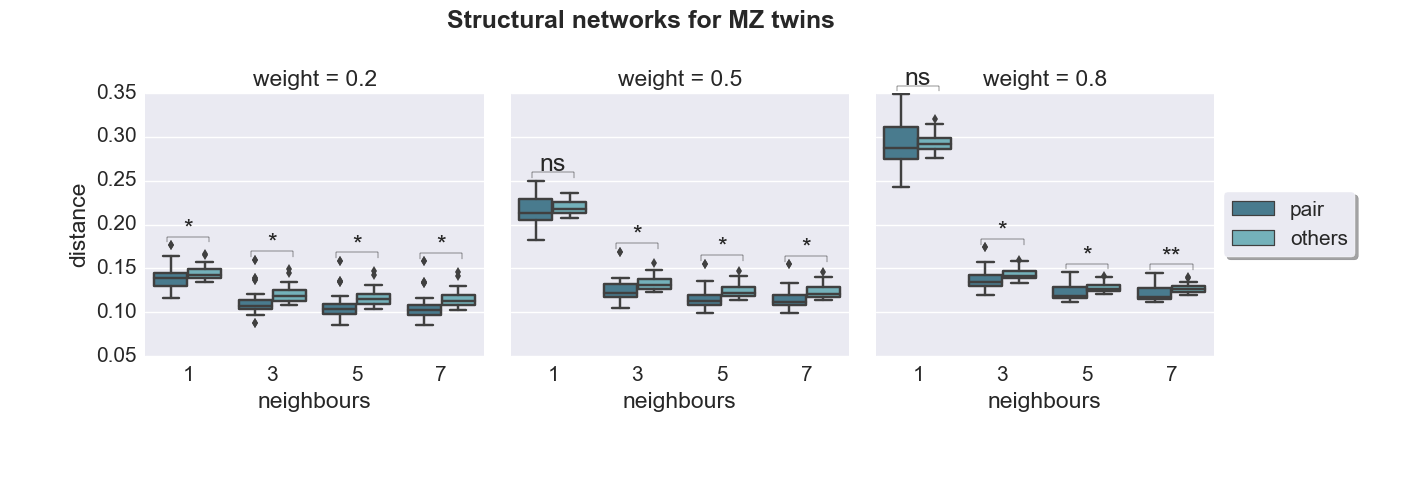} \\
\includegraphics[width=.85\linewidth]{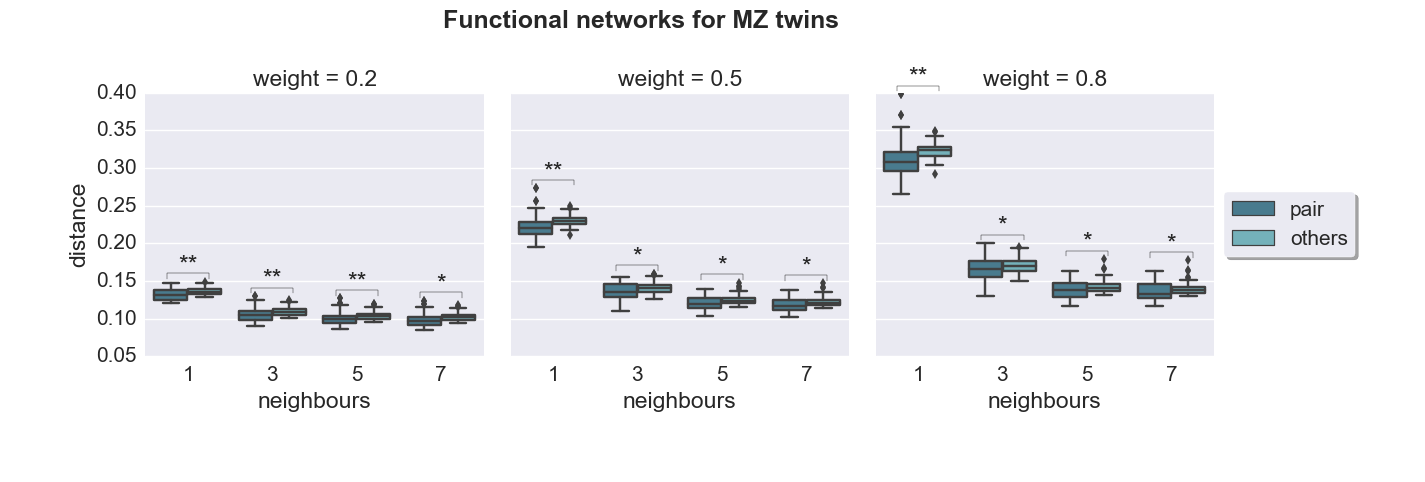} \\
\includegraphics[width=.85\linewidth]{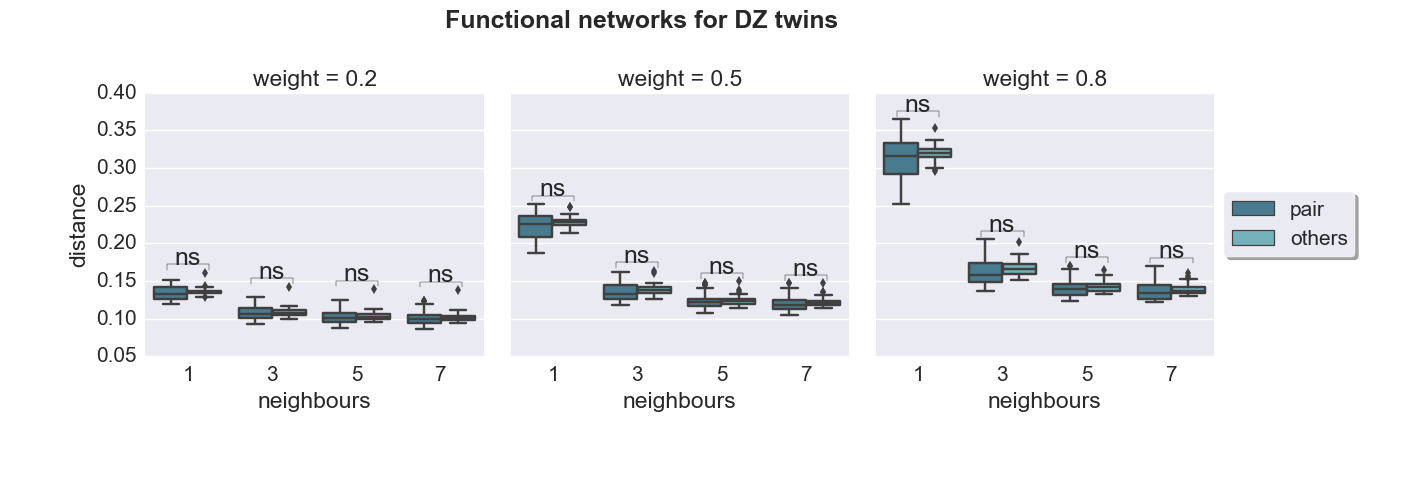}
\caption{Boxplots comparing GED between twin pairs (dark blue) to unrelated pairs (light blue) on structural (top) and functional (middle, bottom) networks derived from single-subject parcellations. All parcellations comprise of 50 parcels.}
\label{fig:MZ_vs_others}
\end{figure}

\subsection{Twin data}
\begin{figure}[t]
\centering
\includegraphics[width=.23\linewidth]{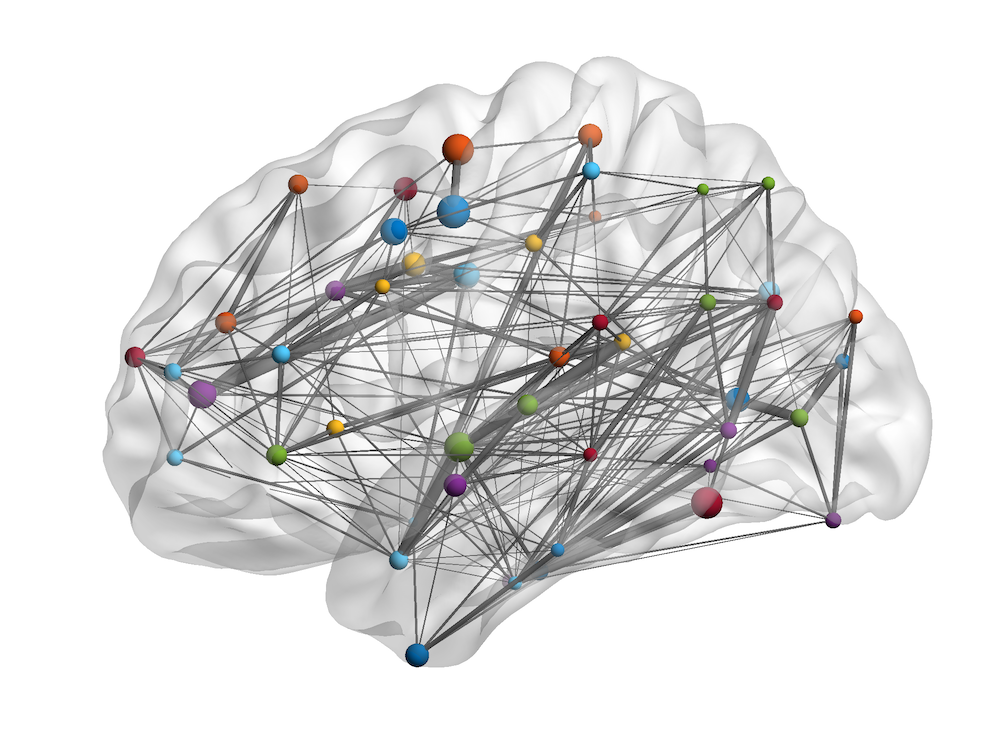}
\includegraphics[width=.23\linewidth]{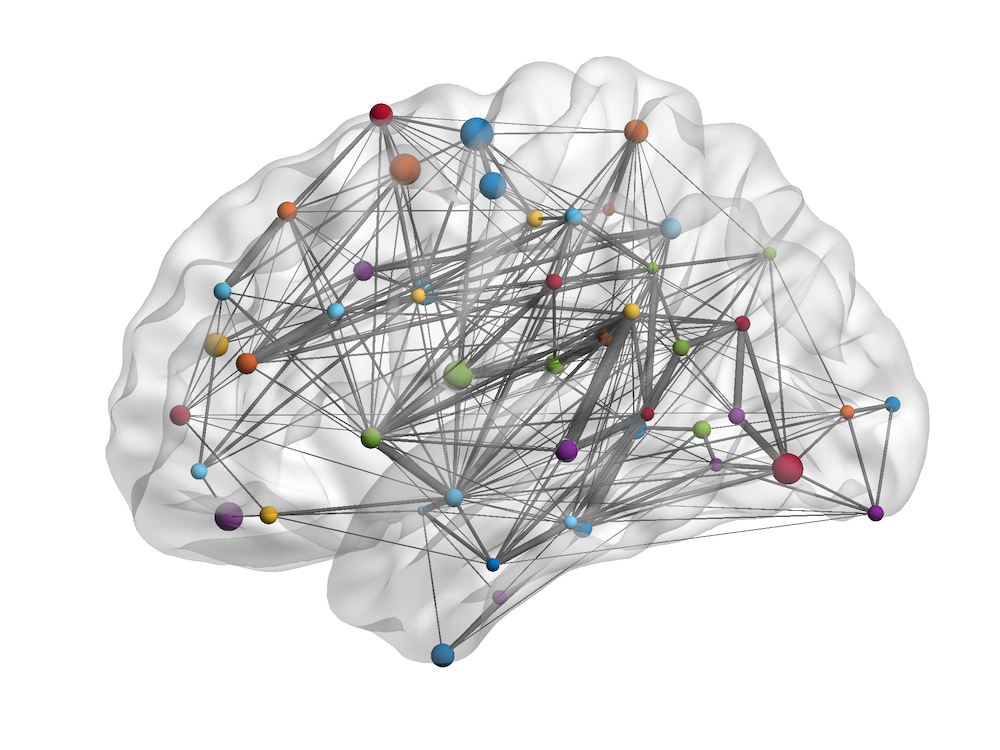}
\hspace{0.5cm}
\includegraphics[width=.23\linewidth]{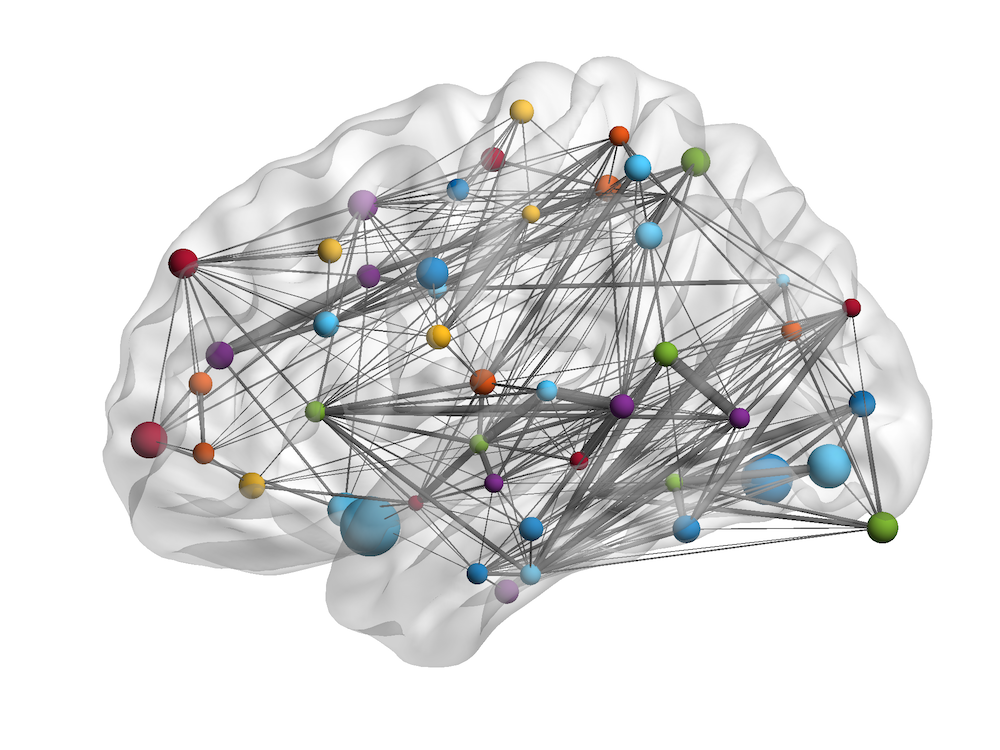}
\includegraphics[width=.23\linewidth]{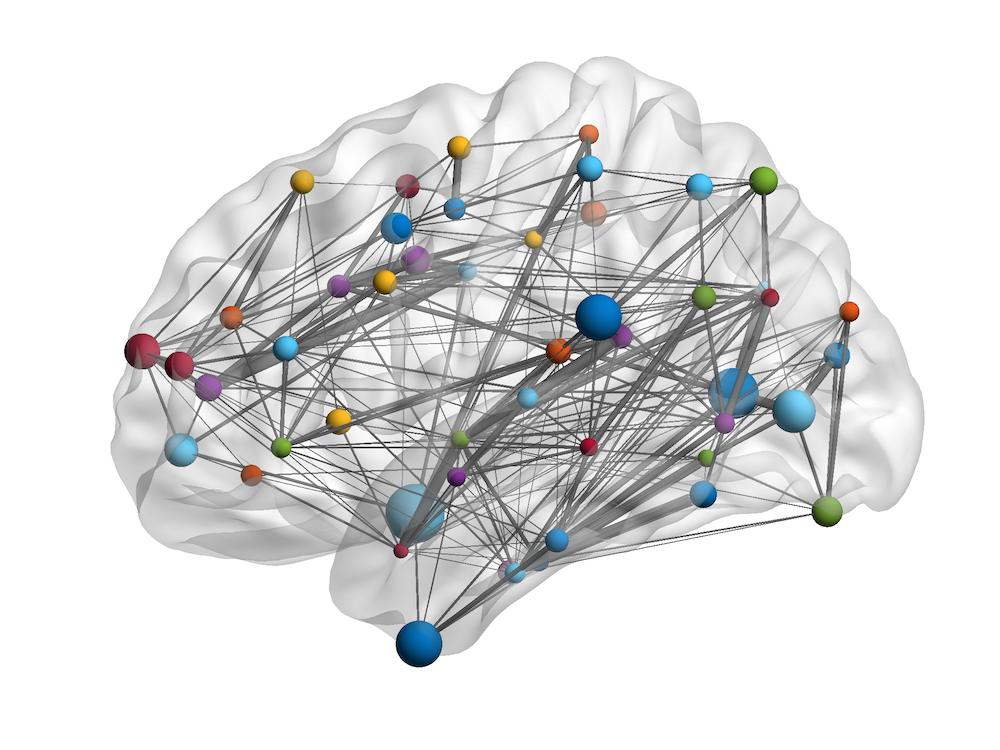}
\\
\includegraphics[width=.23\linewidth]{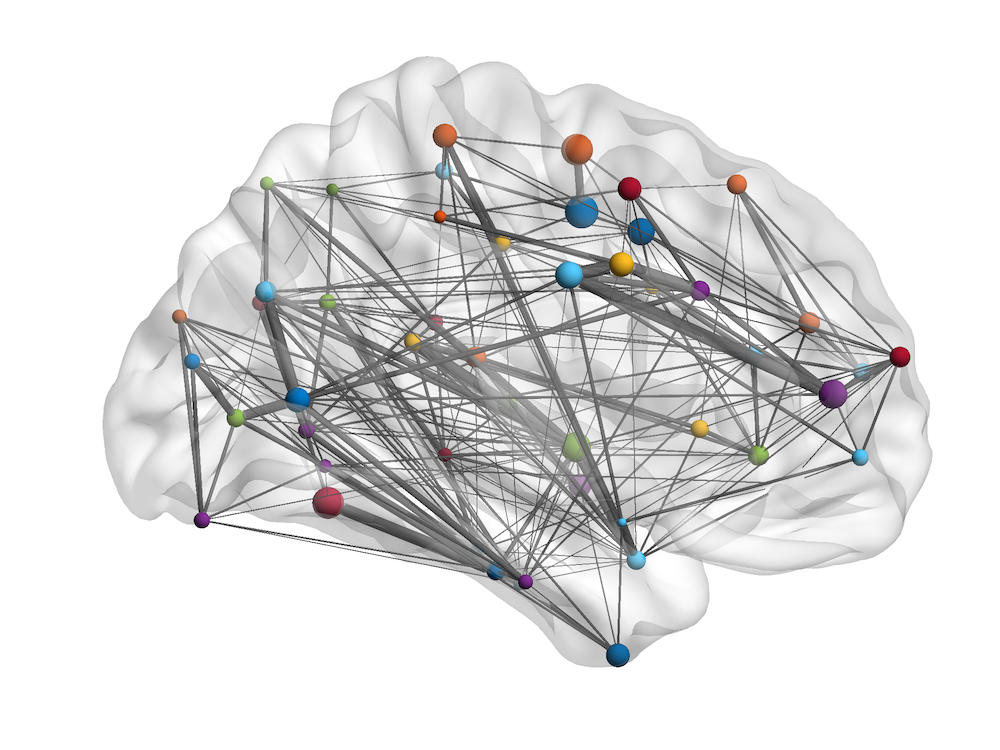}
\includegraphics[width=.23\linewidth]{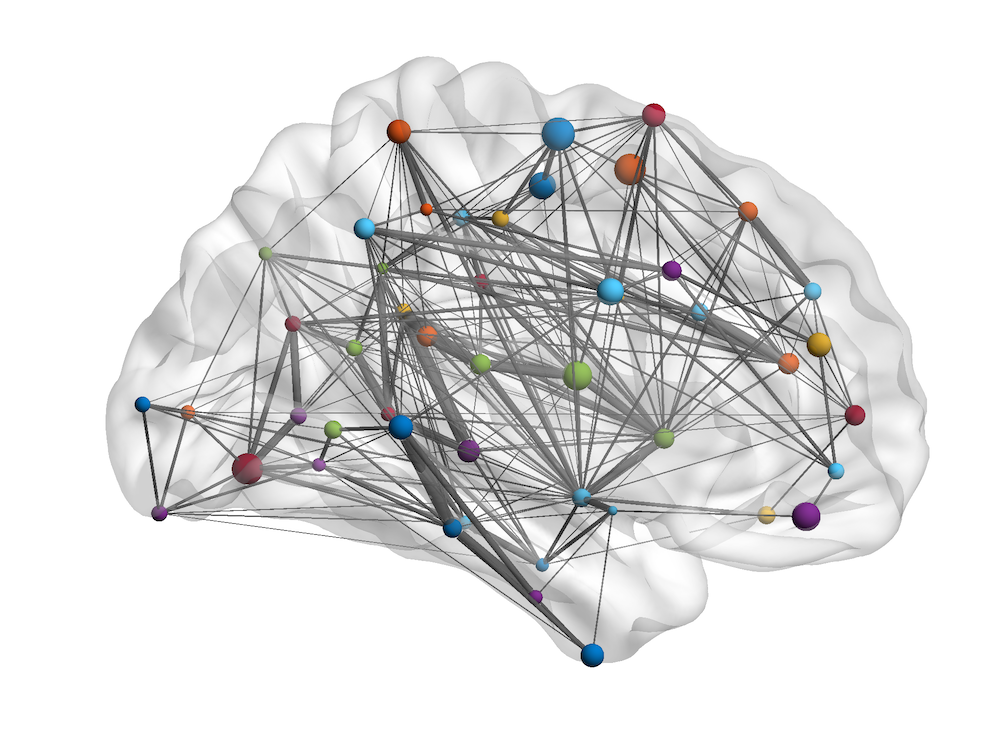}
\hspace{0.5cm}
\includegraphics[width=.23\linewidth]{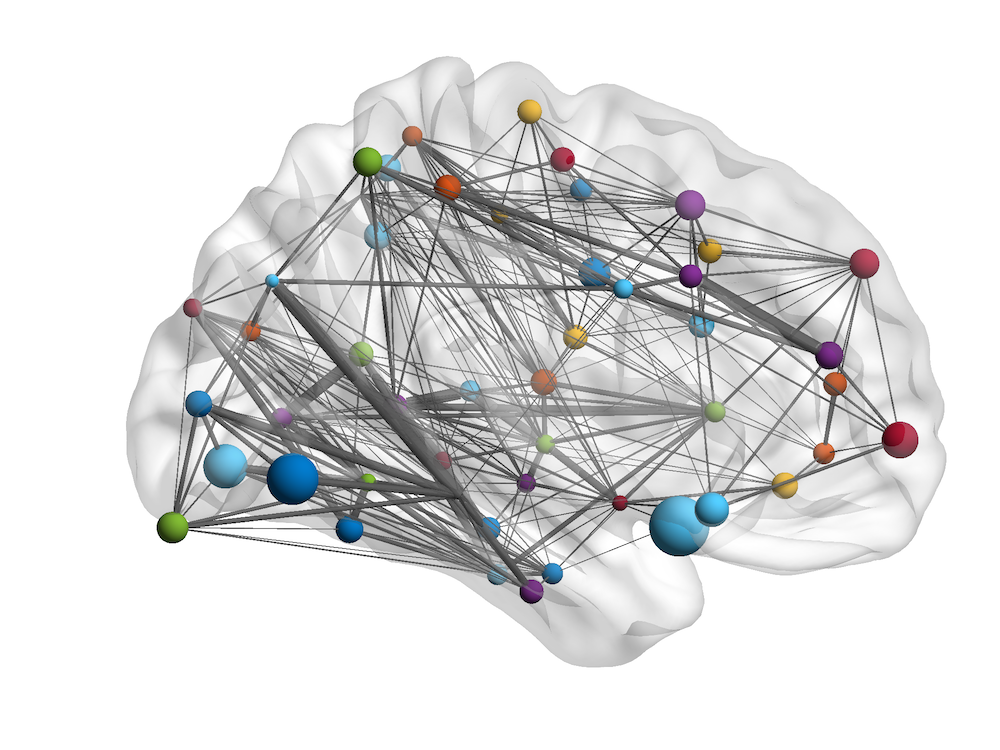}
\includegraphics[width=.23\linewidth]{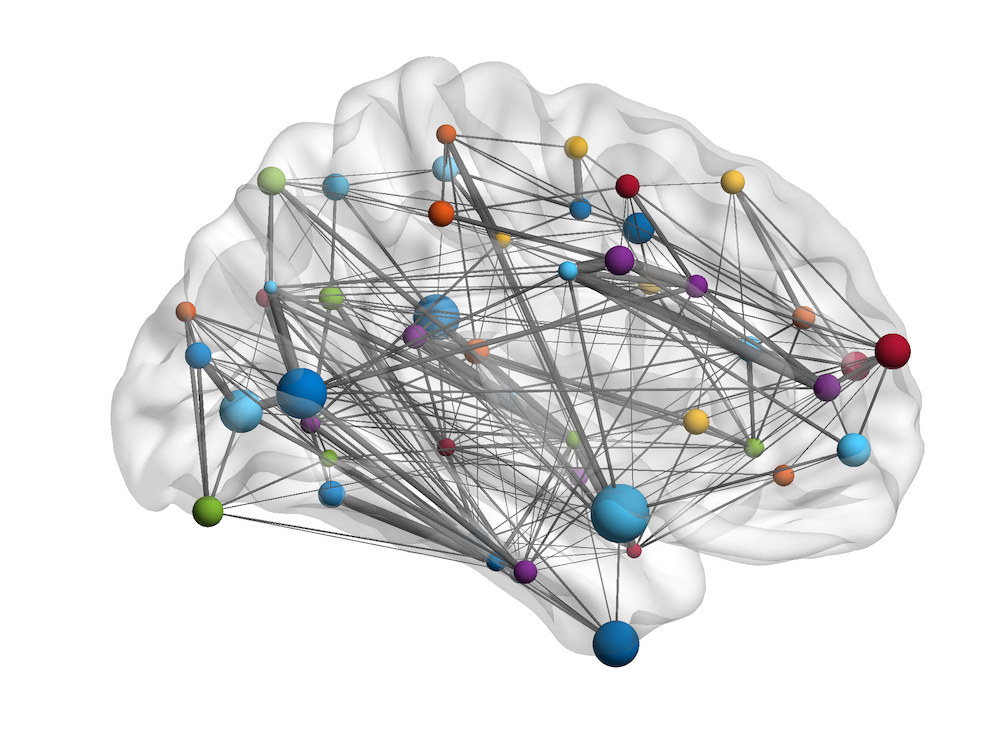}
\caption{Medial and lateral views of a monozygotic twin pair (left) and an unrelated pair (right) with corresponding nodes highlighted with the same color. The size of the nodes indicates the node edit distance.} 
\label{fig:twin_pair}
\end{figure}

In this set of experiments the GED algorithm was first applied to the structural networks of MZ pairs. A permutation test (1000 permutations) was performed between the pair distance and the average distance to unrelated subjects and the results are summarized in Figure~\ref{fig:MZ_vs_others}. It can be observed that GED is significantly discriminative for MZ pairs, but the best separation is achieved for $a \in [0.5, 0.8]$ and $nn = 7$. The algorithm was also applied to the functional networks of both MZ and DZ pairs. The boxplots for MZ twins show that GED yields good separation on functional networks as well (Fig.~\ref{fig:MZ_vs_others}), with the best result in terms of statistical separation achieved for $\alpha=0.2$, while this is not the case for DZ twins. Fig.~\ref{fig:twin_pair} shows an example MZ pair in comparison to an unrelated pair of subjects. It can be observed that node distances are much smaller for the twin pair in comparison to the unrelated pair.

At this point it should be mentioned that we present the results for $q=50$, but results look similar for higher dimensions.

\section{Discussion}
This work proposes a novel way of evaluating similarity between brain networks based on graph edit distance, which incorporates both structural and semantic information about the networks. Hence, it can be applied to networks with different sets or even different number of nodes. Additionally, it enforces spatial constraints to the explored edit operations, reducing the computational time required for the distance calculation. Our approach was applied on healthy subjects and twin pairs and was able to reflect similarities between networks representing the same subject as well as monozygotic twin pairs that share the same genome.

As an extension of this work, different weights can be assigned to each network feature according to their contribution to the task at hand. To provide a complete framework for network comparison independent of the kind of network or the application, the optimal parameters can then be chosen in a cross-validation setting. Additionally, the node correspondences which are provided by this algorithm along with the distance estimate can be used to explore network dynamics in early brain development or neurodegenerative diseases.

\bibliographystyle{splncs}
\bibliography{mymiccaipaper}

\end{document}